\documentclass[
    ,final            
 ]
  {aipproc}

\layoutstyle{8x11single}

\newcommand{\gd}{\dot{\gamma}}
\newcommand{\kap}{\bf{\kappa}}
\newcommand{\rb}{{\bf r}}
\def\gtrapprox{\raise2.5pt\hbox{$>$}\llap{\lower2.5pt\hbox{$\approx$}}}

\begin{document}

\title{Theory of Thermodynamic Stresses in Colloidal Dispersions
at the Glass Transition}


\author{D. Hajnal}{
  address={Fachbereich Physik, Universit\"at Konstanz,
 78457 Konstanz, Germany}
}
\author{O. Henrich}{
  address={Fachbereich Physik, Universit\"at Konstanz,
 78457 Konstanz, Germany}
}
\author{J. J. Crassous}{
  address={Physikalische Chemie I, University of Bayreuth,
95440 Bayreuth, Germany}
}
\author{M. Siebenb{\"u}rger}{
  address={Physikalische Chemie I, University of Bayreuth,
95440 Bayreuth, Germany}
}
\author{M. Drechsler}{
  address={Physikalische Chemie I, University of Bayreuth,
95440 Bayreuth, Germany}
}
\author{M. Ballauff}{
  address={Physikalische Chemie I, University of Bayreuth,
95440 Bayreuth, Germany}
}
\author{M. Fuchs}{
  address={Fachbereich Physik, Universit\"at Konstanz,
 78457 Konstanz, Germany}
}

\begin{abstract}
We discuss the nonlinear rheology of dense colloidal dispersions at
the glass transition. A first principles approach starting with
interacting Brownian particles in given arbitrary homogeneous
(incompressible) flow neglecting hydrodynamic interactions is
sketched. It e.g. explains steady state flow curves for finite shear
rates measured in dense suspensions of thermosensitive core-shell
particles consisting of a polystyrene core and a crosslinked
poly(N-isopropylacrylamide)(PNIPAM) shell. The exponents of simple
and generalized Herschel Bulkley laws are computed for hard spheres.
\end{abstract}

\maketitle


\section{Introduction}

\indent

Soft materials, such as particle dispersions, exhibit a wide range
of rheological properties. While dilute colloids flow with a
viscosity only slightly higher than that of the solvent,
concentrated dispersions behave as soft amorphous solids.  Under
large deformations, they may yield and flow with a strongly
shear-rate dependent viscosity, while their structure can remain
amorphous and homogeneous \cite{Pet:02b}.

In order to address the nonlinear rheology of such viscoelastic
colloidal dispersions, recently, a first-principles constitutive
equation  under arbitrary time-dependent homogeneous flow was
developed \cite{Bra:07}. It includes transitions to  arrested glassy
states as  function of the thermodynamic control parameters, thus
allowing the interaction between slow structural relaxation and
time-dependent external flow to be investigated. The approach
generalizes the integrations through transients (ITT) approach
developed for steadily sheared dispersions \cite{Fuc:02} to the
general situation. It concentrates on the thermodynamic stresses
arising from potential and Brownian interactions between the
particles, because they overwhelm hydrodynamic contributions close
to vitrification and at low deformation rates, and because they are
dominated by slow structural rearrangements, which can be
approximated using mode coupling techniques \cite{Goe:92}.

A colloidal model system of hard spheres was developed
\cite{Cra:06}, and experiments at high densities were performed
\cite{Cra:08} to test the connection between glassy dynamics and
nonlinear rheology, which is at the heart of the theoretical
approach. To this end, linear frequency dependent shear moduli and nonlinear shear rate dependent flow curves were measured and analysed in parallel \cite{Cra:08}. Here, we summarize some of the observations on the steady
state flow curves and provide some new insights into the flow curves
close to the glass transition.

\section{Theory}

\indent

The microscopic theoretical approach \cite{Bra:07} starts with a
system of $N$ spherical Brownian particles interacting via internal
forces and dispersed in a solvent with a specified velocity profile
${\bf v}(\rb,t)=\kap(t)\cdot\rb$. The time-dependent velocity
gradient tensor $\kap(t)$ is assumed spatially constant, thus
excluding the non-homogeneous flows which occur in shear-banded and
shear-localized states. While hydrodynamic interactions and
fluctuations in the velocity profile are thus neglected from the
outset, this Statistical Mechanics model  can be investigated by
Brownian dynamics simulations  \cite{Str:99}, and constitutes a
first microscopic approach to real glassy colloidal suspensions. The
properties of this microscopic model have been worked out for low
densities (partially even including hydrodynamic interactions
\cite{Ber:01c}), and it provides the starting point for various
(approximate) theories for intermediate concentrations.

Considering, for the sake of discussion, the example of steady
shearing \cite{Fuc:02}, the aim of the microscopic approach is to compute the
stationary many-body distribution function, which is possible albeit
only approximately. It describes the complete  probability
distribution of the particle positions, and using it any stationary
average can be computed. Its computation is achieved using formally
exact projection operator manipulations and generalized Green-Kubo
relations, followed by mode coupling approximations, which share
similarities with mean field approximations \cite{Fuc:02}. The
approximations implement the assumption that slow structural
rearrangements dominate the thermodynamic stresses in dense
dispersions, and that they can be approximated using (transient)
density correlation functions.  The later are taken from equations
of motions closed with (again) mode coupling approximations.
Competition of structural slowing down (called 'cage effect') with
flow induced decorrelation ('flow advection' entering via
Cauchy-Green tensors \cite{Bra:07}) and cut-off of memory, determines the
distorted microstructure and the stresses. Universal aspects of the predicted rheology can be captured in 'schematic models', like the $F_{12}^{(\gd)}$-model, which retain the nontrivial dynamical scaling laws of the microscopic approach, but simplify its spatial structure. 

\section{Experimental system}

\indent

The particles consist of a solid core of poly(styrene) onto which a
network of crosslinked poly(N-isopropylacrylamide) (PNIPAM) is
affixed. Immersed in water the shell swells at low temperatures.
Raising the temperature above 32$^o$C leads to a collapse within the
shell. Screening the remaining electrostatic interactions by adding
salt, the system crystallises such that its phase diagram can be
mapped onto the one of hard spheres\cite{Cra:06}. The dependence of
the packing fraction  $\phi_{\rm eff}$ (viz.~the relative volume
taken by the spheres) on the temperature is given by the
hydrodynamic radius $R_H$ determined from dynamic light scattering
in the dilute regime. The flow curves were measured with a
stress-controlled rotational rheometer MCR 301 (Anton Paar),
equipped with a Searle system \cite{Cra:08}, upon decreasing the
shear rate $\dot{\gamma}$ and checking carefully for stationarity.
\begin{figure}
  \includegraphics[height=.35\textheight]{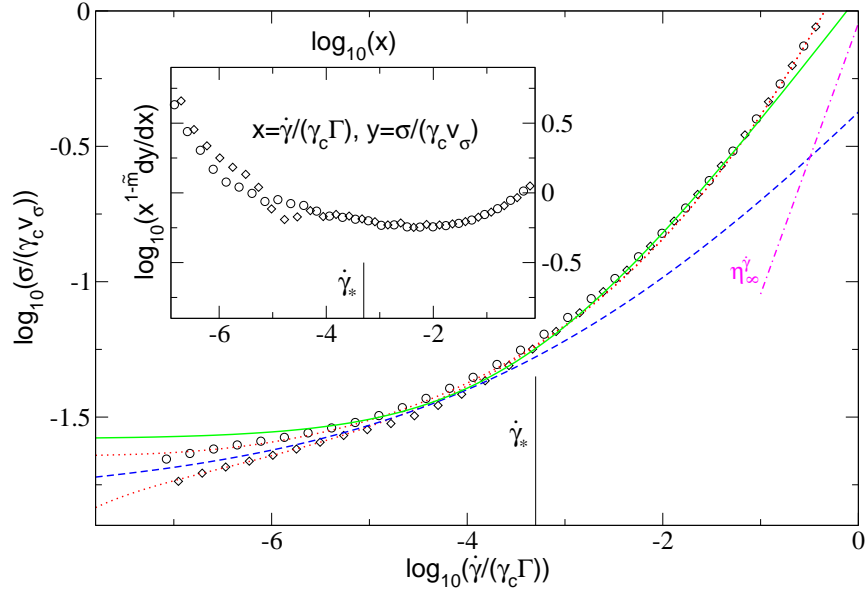}
\caption{(Color online) Flow curves of a  suspension of colloidal
hard spheres for two densities close to the glass transition,
$\phi_{\rm eff}=0.580$ (fluid state, diamonds) and $\phi_{\rm
eff}=0.608$ (glassy state, circles) rescaled to collapse at high
shear; from Ref.~\cite{Cra:08}. Parameters are $\Gamma=60D_{0}/R_{H}^{2}$, $\gamma_c=0.1$ and $v_\sigma=50k_{B}T/R_{H}^{3}$
for $\phi_{\rm eff}=0.580$, and
$\Gamma=84D_{0}/R_{H}^{2}$, $\gamma_c=0.105$ and $v_\sigma=77k_{B}T/R_{H}^{3}$ for $\phi_{\rm eff}=0.608$, respectively, where $R_H$ is the hydrodynamic radius and $D_0$ the single particle diffusion coefficient at infinite dilution; $k_BT$ is the thermal energy.  Fits
using the F$_{12}^{(\gd)}$-model are given as dotted red  lines
\cite{Cra:08}. The blue dashed cuves gives the generalized
Herschel-Bulkley law Eq.~(\ref{eq1}) with $m=0.143$, which holds
asymptotically for $\gd\ll\gd_*$ at the glass transition;
$\sigma^+_c=0.10\, k_BT/R_H^3$ is the dynamic yield stress at the transition.  The
green solid curve is the Herschel-Bulkley law (\ref{eq2}) which
holds for intermediate shear rates $\gd_*\ll\gd\ll\Gamma$ at the
transition; the exponent $\tilde m=0.473$ follows for hard spheres
using the Percus-Yevick approximation for the equilibrium structure
factor \cite{Haj:08,Goe:92}. The magenta dot-dashed straight line
labeled $\eta_\infty^{\gd}$ denotes the (extrapolated) linear
high-shear asymptote. The inset tests the exponent $\tilde m$, as
$d\sigma/d\gd \times \gd^{1-\tilde m}\propto {\rm const.}$ should
hold for the Herschel-Bulkley law.  \label{Fighb}}
\end{figure}

\section{Comparison of theory and experiment}

\indent

The  microscopic ITT equations contain a non-equilibrium transition
between a fluid and a shear-molten glassy state. Close to the
transition, (rather) universal predictions can be made about the
non-linear dispersion rheology and the steady state properties. A
qualitative change in the flow curves is predicted, where the
stationary stress exhibits a Newtonian viscosity
$\sigma(\gd\to0)\to\eta_0 \gd$ in fluid states, while it exhibits a
dynamic yield stress $\sigma(\gd\to0)\to\sigma^+$ in glassy ones.
Non-trivial power-laws in the flow curves exist close to the glass
transition itself; for hard spheres, it is estimated to lie at $\phi_{\rm eff}=0.580$
\cite{Cra:08}. Right at the glass transition, a generalized
Herschel-Bulkley law
\begin{equation}\label{eq1}
\sigma(\gd\to0) \to \sigma^+_c \left( 1 + |{\gd}/{\gd_*}|^m + c_2 \,
|{\gd}/{\gd_*}|^{2m} + c_3 \, |{\gd}/{\gd_*}|^{3m } \right)
\end{equation}
holds and describes the flow curves over an appreciable part of the
range $\gd\le\gd_*$; the exponent is $m=0.151$ for the exponent
parameter $\lambda=0.735$ computed within mode coupling theory for
hard spheres using the Percus-Yevick approximation.

Beyond the asymptotic regime, a (simple) Herschel-Bulkley law
appears in the flow curves  of the F$_{12}^{(\gd)}$-model \cite{Haj:08}
\begin{equation}\label{eq2}
\sigma(\gd \gtrapprox \gd_* ) = \tilde \sigma_0 + \tilde \sigma_1
|\gd|^{\tilde m} \;.
\end{equation}
 Its exponent can again be calculated from $\lambda$, which
for hard spheres gives  $\tilde m=0.473$. This law provides a
semi-quantitative fit  of the flow curves for around two decades in
$\gd$ close to the glass transition as shown in Fig.~\ref{Fighb}.

The dynamic yield stress $\sigma^+$ can be read off by extrapolating
the flow curve in the glass to vanishing shear rate. While this
agreement supports the prediction of a dynamic yield stress in the
ITT approach and demonstrates the usefulness of this concept, small
deviations in the flow curve at low $\gd$ are present. They indicate
the existence of an additional slow dissipative process which may
correspond to a melting of the glass at even longer times or smaller
shear rates beyond the experimental window.

\section{Conclusions and outlook}

\indent

Recently, the connection was emphasized between the physics of the
glass transition and the rheology of dense colloidal dispersions,
including in strong homogeneous flow \cite{Bra:07,Fuc:02}. The
equilibrium structure factor and one time scale (obtained from
matching to short time motion and affected by hydrodynamic
interactions) are required as inputs to the derived constitutive
equation. Using model colloidal particles made of thermosensitive
core-shell particles the predictions could be investigated in the
vicinity of the transition between two stationary states under
steady shear, a shear-thinning fluid and a shear-molten glass
\cite{Cra:06,Cra:08}.


\begin{theacknowledgments}
 We  acknowledge
financial support by the Deutsche Forschungsgemeinschaft in SFB TR6,
SFB 481, and SPP 608.
\end{theacknowledgments}

\end{document}